\documentclass[12pt]{iopart}

\usepackage{url}  
\usepackage{graphicx}  
\usepackage{atlasphysics}
\usepackage{lineno}
\begin{document}

\title[Recent Heavy Ion Results with the ATLAS Detector at the LHC]
{Recent Heavy Ion Results with the ATLAS Detector at the LHC}
\author{Peter Steinberg, for the ATLAS Collaboration}
\address{Physics Department, Brookhaven National Laboratory, Upton, NY 11973, USA}
\ead{peter.steinberg@bnl.gov}
\begin{abstract}
Results are presented from the ATLAS collaboration from the 2010 LHC heavy ion run,
during which nearly 10 inverse microbarns of luminosity were delivered.  
Soft physics results include charged particle multiplicities and 
collective flow.
The charged particle multiplicity, which tracks initial state entropy production,
increases by a factor of two relative to the top RHIC energy, with a 
centrality dependence very similar to that already measured at RHIC.
Measurements of elliptic flow out to large transverse
momentum also show similar results to what was measured at RHIC,
but no significant pseudorapidity dependence.
Extensions of these measurements to higher harmonics have also been made,
and can be used to explain structures in the two-particle correlation 
functions that had long been attributed to jet-medium interactions.
New hard probe measurements include single muons, jets and high $p_T$ hadrons.
Single muons at high momentum are used to extract the yield of $W^{\pm}$ bosons
and are found to be consistent within statistical uncertainties 
with binary collision scaling.
Conversely, jets are found to be suppressed in central events by a factor of
two relative to peripheral events, 
with no significant dependence on the jet energy.  Fragmentation functions
are also found to be the same in central and peripheral events.
Finally, charged hadrons have been measured out to 30 GeV, and their centrality
dependence relative to peripheral events is similar to that found for jets.

\end{abstract}

\section{Heavy ions with the ATLAS detector}
\begin{figure}[t]
\begin{center}
\includegraphics[width=19pc]{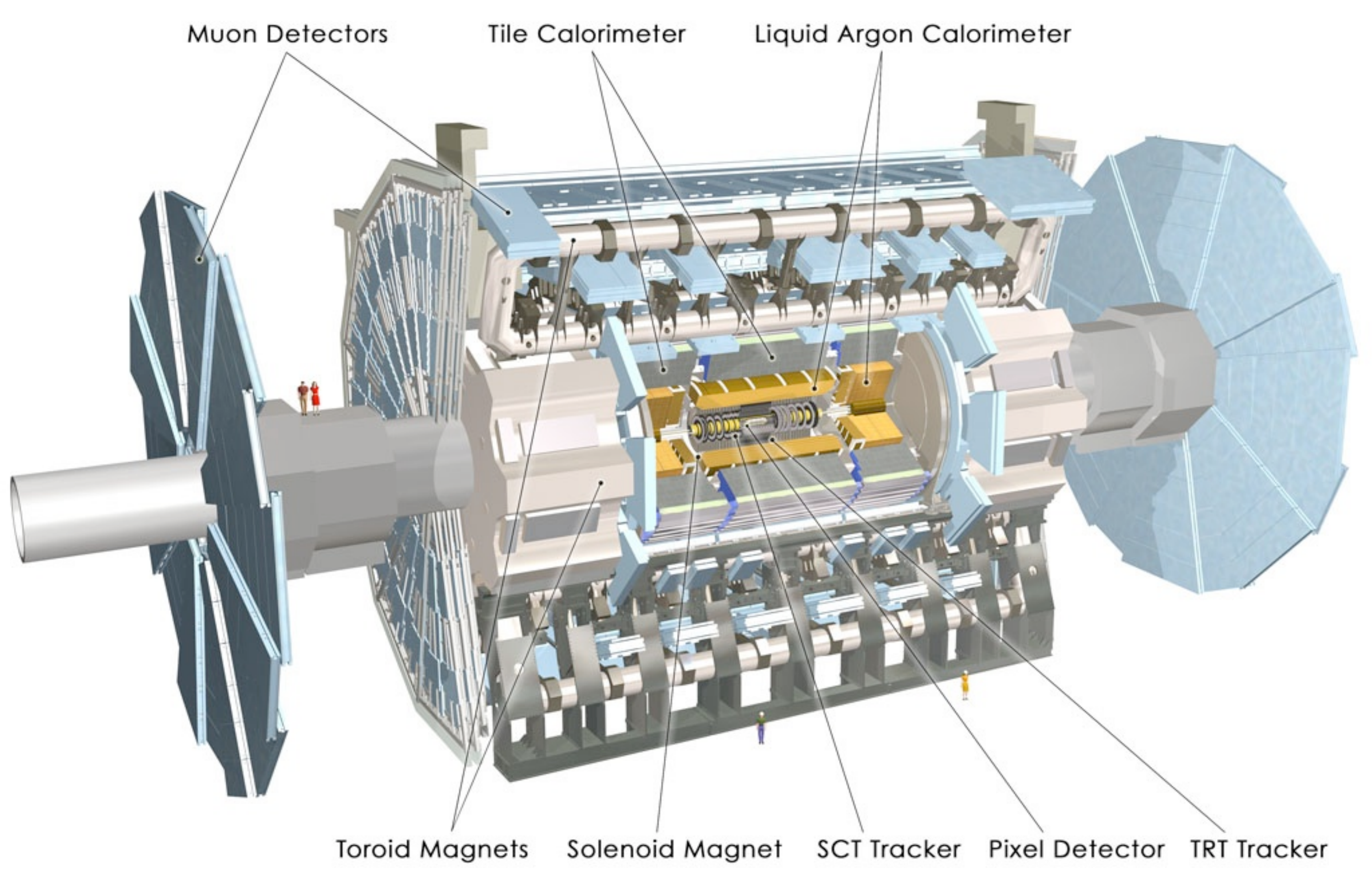}
\includegraphics[width=16pc]{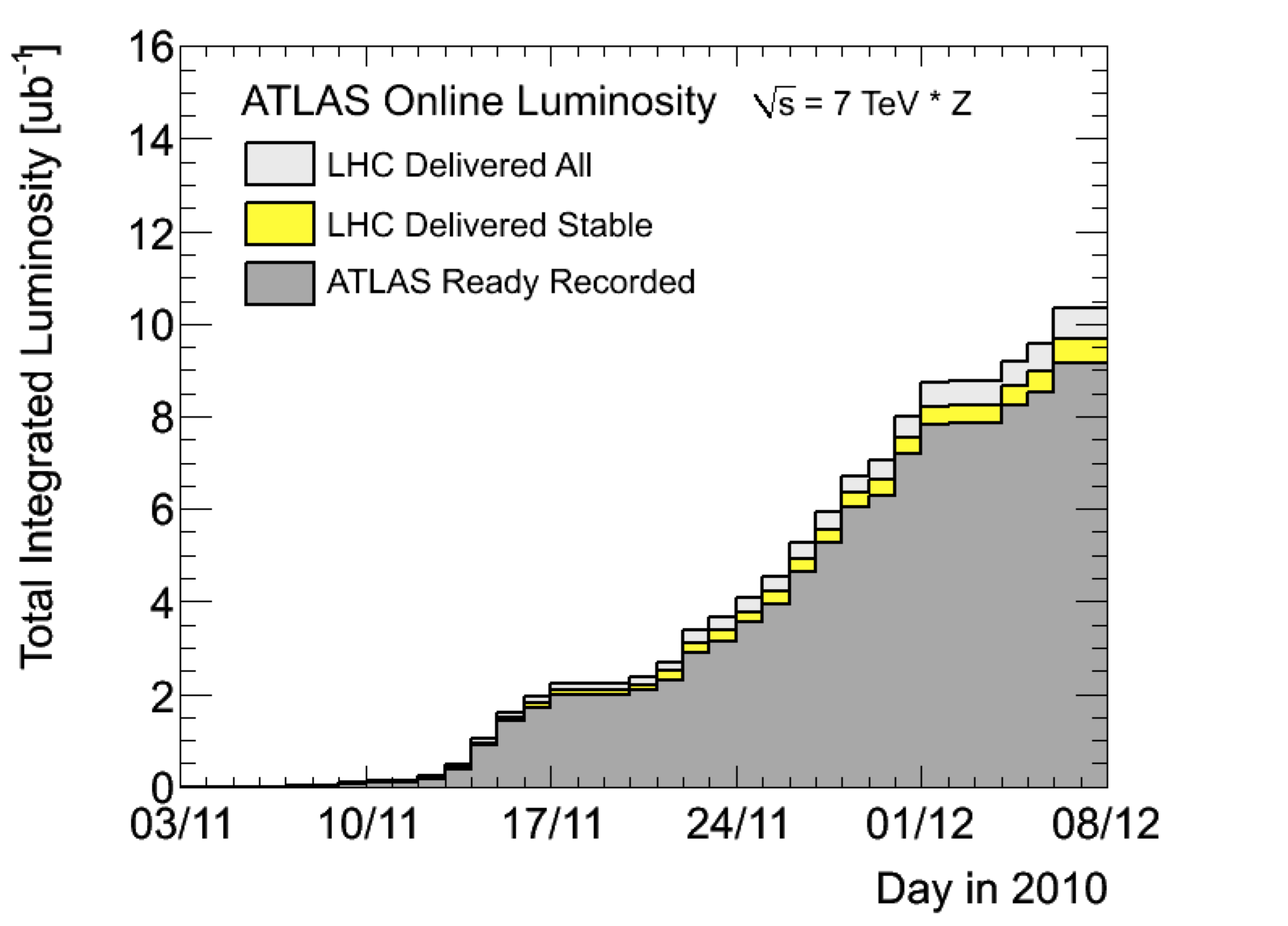}
\caption{\label{atlas}
(left) Schematic diagram of the ATLAS detector, showing the three main subsystems: the\
 Inner Detector ($|\eta|<2.5$),
the calorimeter ($|\eta|<4.9$) and the muon spectrometer ($|\eta|<2.7$).
(right) Integrated luminosity taken by ATLAS in the 2010 heavy ion run vs. time.
}
\end{center}
\end{figure}

The ATLAS detector at the LHC~\cite{Aad:2008zzm}, shown in Fig.~\ref{atlas}, 
was designed primarily for precise measurements
in the highest-energy proton-proton collisions, particularly to discover
new high-mass particle states.  However, it is also a very capable detector
for the measurement of heavy ion collisions at the highest energies achieved
to date, already a factor of 14 higher than available at the Relativistic
Heavy Ion Collider.
The ATLAS Inner Detector is immersed in a 2 Tesla solenoid magnetic field and
provides precise reconstruction of particle trajectories typically
with three pixel layers and four double sided silicon strip detectors
out to $|\eta|=2.5$, as
well as a transition radiation tracker, out to $|\eta|=2$.
The longitudinally segmented ATLAS calorimeter provides electromagnetic
and hadronic measurements out to $|\eta|=4.9$, with particularly
high spatial precision in the $\eta$ direction.  
Finally, the ATLAS Muon Spectrometer is located outside the calorimeters
(which range out most of the hadronic backgrounds) and measures muon
tracks out to $|\eta|<2.7$.

The LHC provided lead-lead collisions to the three large experiments in
November and December 2010, and ATLAS accumulated almost ten inverse microbarns of luminosity,
as shown in the right panel of Fig.~\ref{atlas}.
The analyses shown here typically use up to eight inverse microbarns,
during which the main solenoid was activated.  A smaller dataset with
field off was used for multiplicity analyses.

\section{Global observables}

\begin{figure}[t]
\begin{center}
\includegraphics[width=18pc]{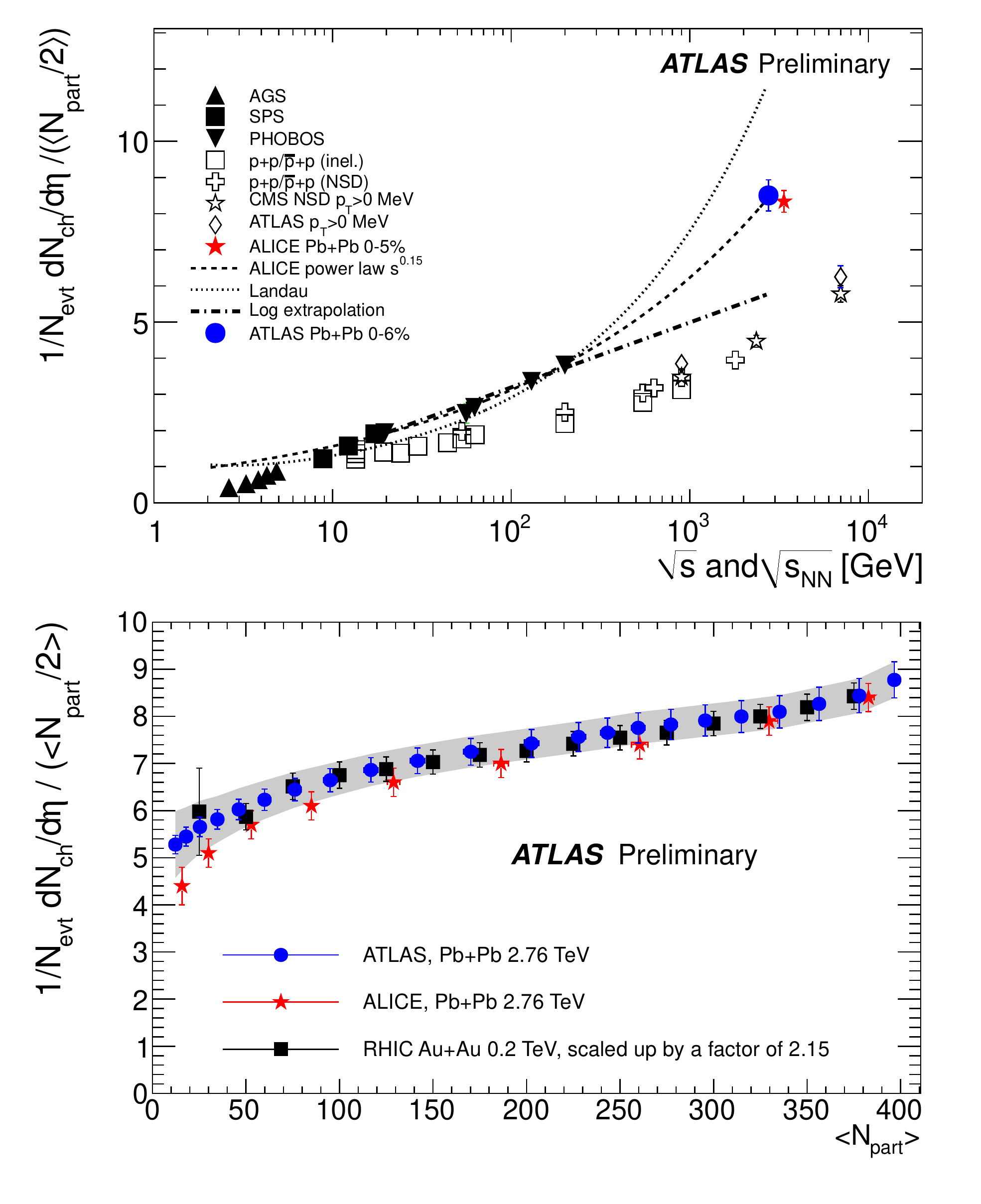}
\caption{\label{paper_dndetaNpart}
(top) Inclusive charged particle multiplicity per participant pair,
measured for the 6\% most central collisions, compared with the earlier
ALICE result, as well as a compilation of lower energy data.
(bottom) The centrality dependence of the multiplicity per participant
pair, compared with ALICE data and an average of the RHIC data.
}
\end{center}
\end{figure}

Global observables, such as the total multiplicity and elliptic flow,
address crucial issues related to the basic
properties of the hot, dense medium.  
The charged particle multiplicity is important both from a thermodynamic
standpoint, as it is proportional to the initial state entropy, as well
as from the perspective of energy loss calculations which require an
estimate of the initial gluon density.
In ATLAS, the multiplicity has been measured~\cite{QM2011Mult} with the solenoid field turned
off to mitigate the losses from the bending of low momentum particles.
The pixel layers within $|\eta|<2$ are used to minimize the non-primary
backgrounds by only using hits consistent with a minimum-ionizing
particle emanating from
the primary vertex.  The pixel hits are assembled both into ``tracklets'' (two
hits, both consistent with the measured primary vertex) and full
tracks (three points, reconstructed with the main ATLAS tracking software).
In all, three methods are used, each with quite different systematic uncertainties,
providing accurate measurements up to the highest particle densities
and uncertainties of only a few percent.
The collision centrality is estimated using the total transverse energy
measured in the ATLAS Forward Calorimeter (FCal,
covering $3.2 < |\eta| < 4.9$) and bins are selected 
using an estimated sampling fraction of $f=100\pm2 \%$.

The multiplicity results are shown in Fig.~\ref{paper_dndetaNpart}.
The yield averaged over $|\eta|<0.5$, $dN_{\mathrm{ch}}/d\eta / \langle N_{part}/2\rangle$,
agrees well (consistent within the stated systematic uncertainties) 
with the previously-available ALICE data~\cite{Collaboration:2010cz}, 
both in the most 
central 0-6\% events and as a function of collision centrality (parametrized
by $N_{part}$).  After scaling the RHIC data by the observed factor of
2.15, the centrality dependence
at the LHC is also seen to be quite similar to that 
observed at a much lower energy.

\begin{figure}[t]
\begin{center}
\includegraphics[width=20pc]{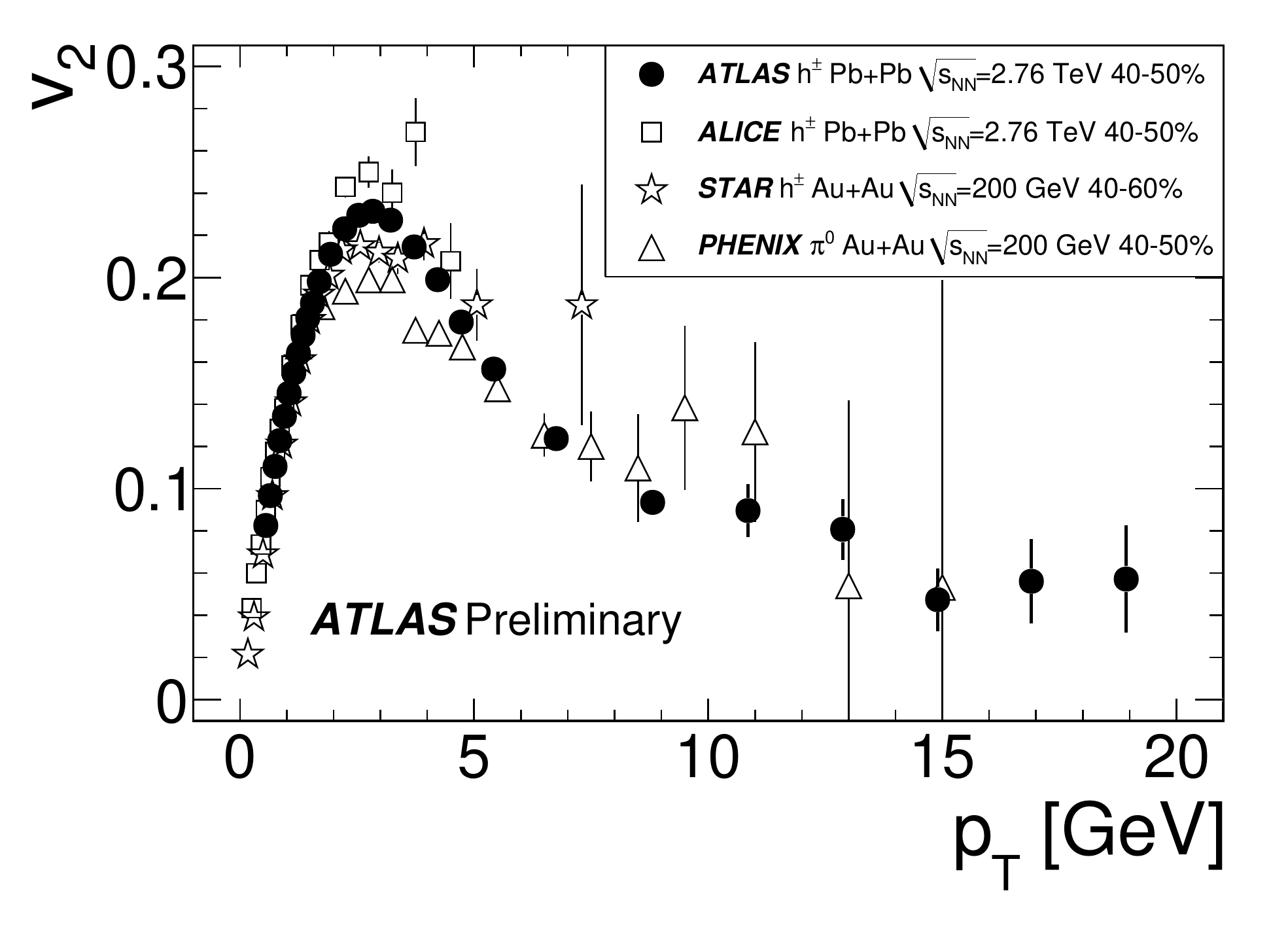}
\caption{\label{v2_vs_pt}
Elliptic flow as a function of transverse momentum for the 40-50\% centrality bin, compared with data from RHIC and the LHC.
}
\end{center}
\end{figure}

The collective response of the system to the initial geometric configuration
of the nucleons in the colliding nuclei is typically characterized by 
the ``elliptic flow'', or a $\cos(2\phi)$ modulation in the 
azimuthal angular distributions
relative to a measured ``event plane.''
In ATLAS, the event plane is measured in the FCal which
subtends $|\eta|=3.2-4.9$.
Elliptic flow is measured using tracks in the Inner Detector 
out to $|\eta|=2.5$.  The effects of ``non flow'' are minimized by
maximizing the gap between the track and the event plane measurement by using
the event plane from the opposite FCal hemisphere 
to the measured track~\cite{QM2011flow}.
The elliptic flow ($v_2$) as a function of transverse momentum and pseudorapidity
has been measured out to $\pT=20$ GeV and $|\eta|<2.5$ for eight centrality bins.
There is no substantial pseudorapidity dependence out to the maximum measured
$\eta$, but there is a strong $\pT$ dependence with a pronounced rise in $v_2$ out
to $\pT=3$ GeV, a slower decrease out to $\pT=8$ GeV and then a very weak
dependence out to the highest measured $\pT$.
Fig.~\ref{v2_vs_pt} compares the ATLAS data with already-published
ALICE~\cite{alicepaper} data as well as lower energy data from PHENIX~\cite{Adare:2010sp} 
and STAR~\cite{Adams:2004bi}, 
all from the 40-50\% centrality bin (except STAR data from 40-60\%).  
One finds that all of the data is quite similar, even at high $\pT$
within the large statistical errors of the PHENIX $\pi^0$ data.
It is not obvious if this apparent scaling behavior is consistent with 
expectations from differential energy loss.

\begin{figure}[t]
\begin{center}
\includegraphics[width=20pc]{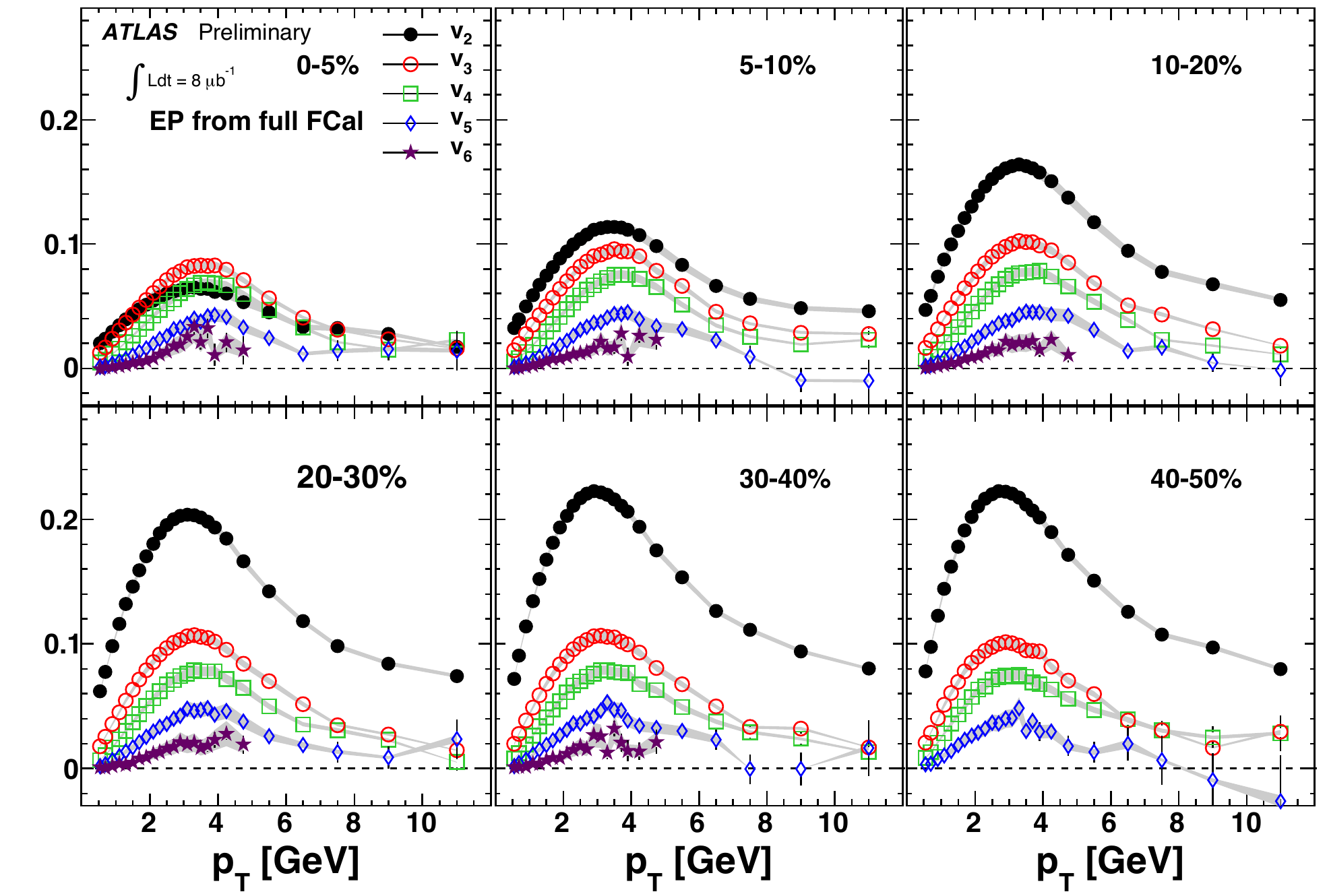}
\caption{\label{vn_vs_pt}
Higher order Fourier coefficients ($v_2-v_6$) as a function of transverse
momentum for six centrality bins.  It is observed that while $v_2$ varies
strongly with centrality, $v_3-v_6$ are nearly invariant.  
}
\end{center}
\end{figure}

After the realization that initial state fluctuations can induce 
higher Fourier modes in the overlap region of the two nuclei~\cite{Alver:2010gr}, elliptic
flow has now been joined by the study of higher-order harmonic
flow coefficients~\cite{QM2011flow}.  These are measured each in their own event plane.
The resolution has been evaluated by a variety of subdetector combinations
and it is found that ATLAS can resolve modes up to $n=6$ in the most
central 40\% of events.
These are shown as a function of $\pT$ in Fig.~\ref{vn_vs_pt} in 
six centrality bins.  One sees the rapid change of $v_2$
with centrality, expected from previous measurements, 
but only a modest change in the higher order
coefficients, suggesting that the latter are mainly sensitive to 
fluctuations of nucleon positions.
Still, it is expected that the information from these higher coefficients
can distinguish between competing physics scenarios attempting to 
explain the initial state 
as well as providing  new information on the viscosity to entropy ratio.


\begin{figure}[t]
\begin{center}
\includegraphics[width=20pc]{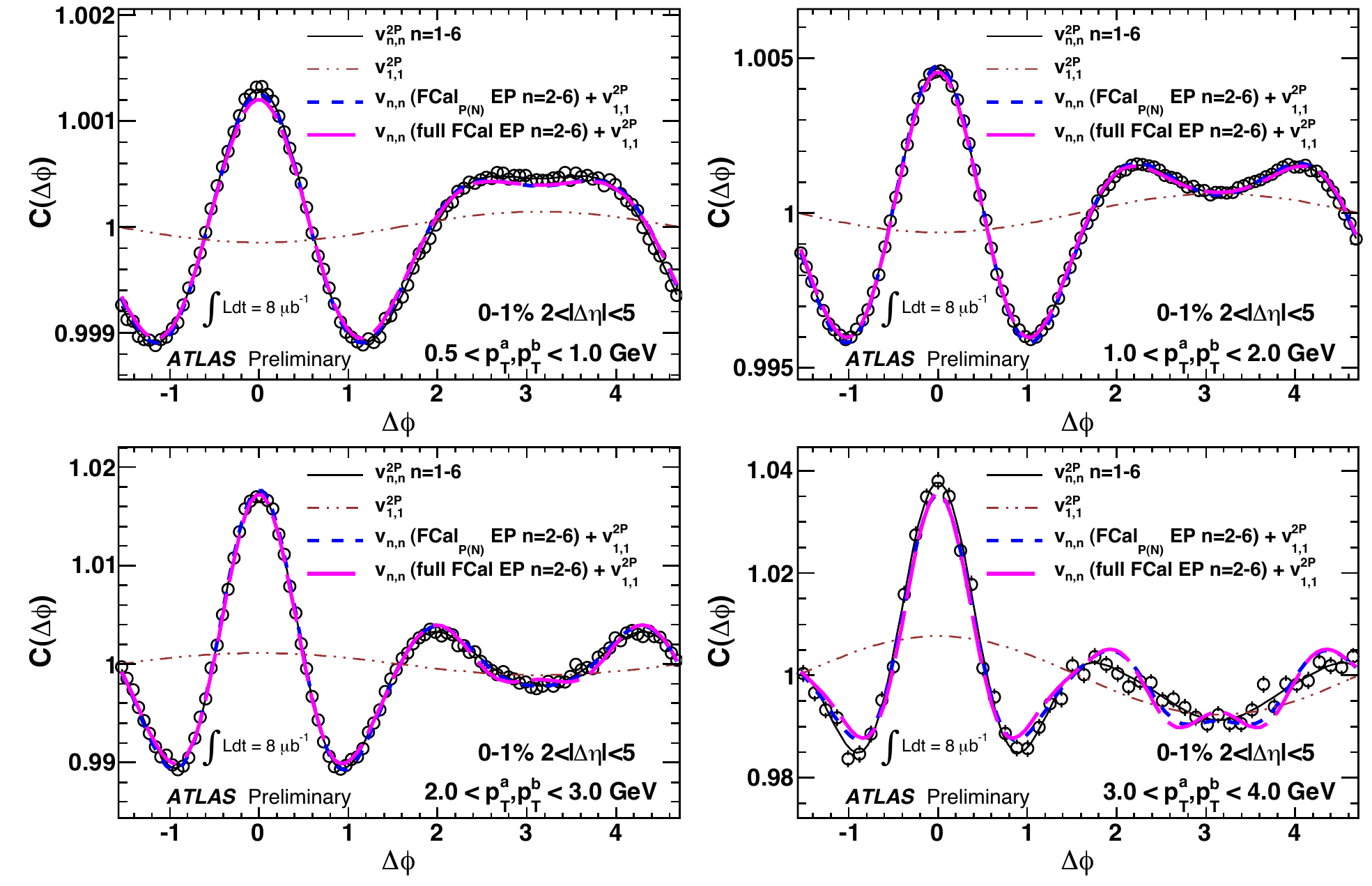}
\caption{\label{2pc}
Measured two-particle correlation function for $2<\pT < 3$ GeV 
and $\eta < 2.5$ in the 1\% most central collisions (open circles).
The reconstructed correlation function using the coefficients from
the event plane measurement (and $v_1$ from the two-particle measurement)
is also shown and agrees quite well with the directly measured correlation
function, using both full FCal (solid line) and FCal$_\mathrm{P(N)}$ (dashed line) methods.
}
\end{center}
\end{figure}

A longstanding puzzle at RHIC is the observation of unusual structures 
(e.g. the Mach Cone~\cite{CasalderreySolana:2004qm})
in the two particle correlation function, particularly on the ``away side''
relative to a high $\pT$ particle.  
Both STAR~\cite{Adams:2005ph} and PHENIX~\cite{Adare:2007vu} observed a strongly-modified away-side 
peak which even developed a ``dip'' at $\phi = \pi$ in the most central events. 
However, this was explained to arise most likely from 
the presence of higher 
flow coefficients beyond elliptic flow~\cite{Alver:2010gr}.
In ATLAS, the correlation function has been measured with large separation in $\eta$ to
suppress the contribution from jets and a discrete Fourier transform (DFT)
is used to extract coefficients out to $n=6$~\cite{QM2011flow}.  These are found to agree
quite well with the $v_n$ extracted using event plane approaches.  More
interestingly, when the event plane coefficients are used to construct
a comparable two particle correlation function $dN/d\phi \propto \sum_n v^2_n \cos(n\phi)$, one finds a striking agreement with the correlations 
measured directly, as shown in Fig.~\ref{2pc}.  
This suggests that what were formerly thought to be indications of
jet-medium interactions may well result simply from the presence of 
higher-order flow harmonics, arising through
a combination of nucleon position fluctuations and viscous effects.

\section{Hard Probes}

\begin{figure}[t]
\begin{center}
\includegraphics[width=14pc]{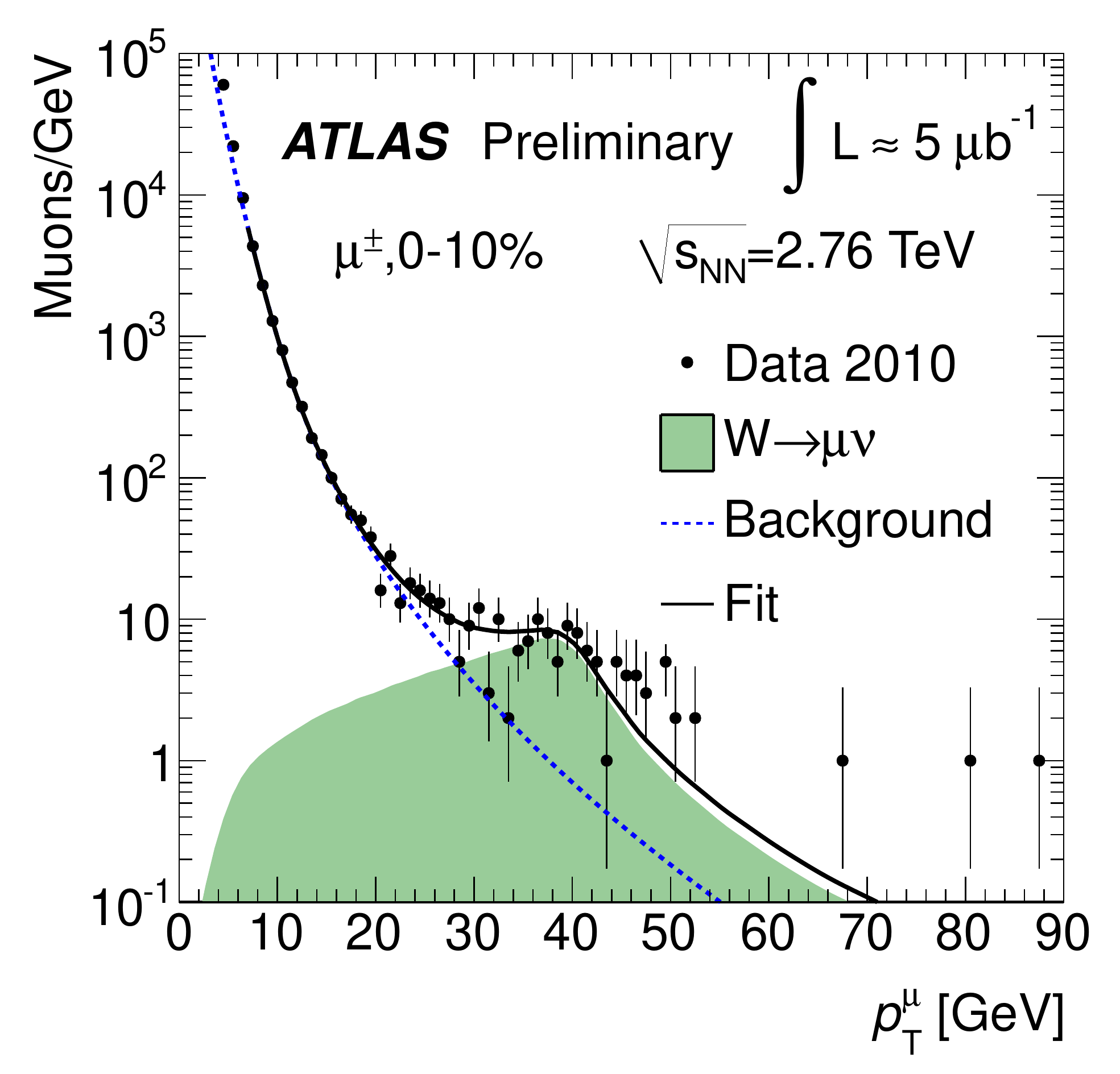}
\includegraphics[width=14pc]{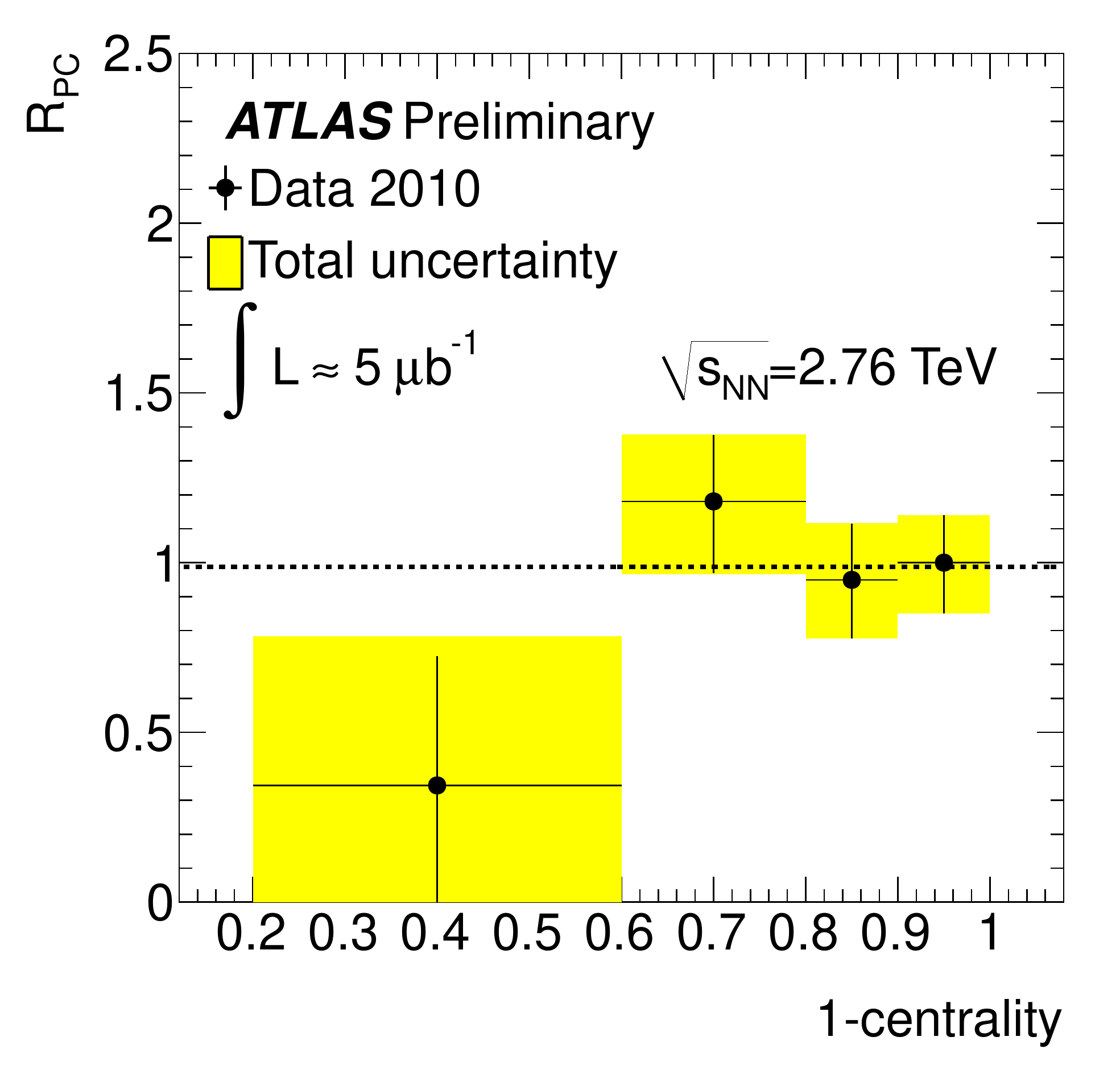}
\caption{\label{W}
(left) Single muon spectrum measured for the 0-10\% most central events,
with the templates for $W$ bosons and heavy flavor (indicated as ``Background'') 
also shown to illustrate the yield extraction procedure.
(right) $R_{PC}$ for $W$ bosons as a function of centrality, showing 
consistency with binary collision scaling.  The dotted line is a fit to
a constant.
}
\end{center}
\end{figure}

High $p_T$ processes produced by perturbative physics in the
first collisions of the two nuclei (also known as ``hard probes'') 
provide a means to probe the conditions
of the collision at very early times.  This is typically done by means of ``suppression''
observables which indicate deviations from the expected scaling of yields
by the number of binary collisions, estimated using Glauber modeling
(e.g. the ATLAS $J/\psi$ measurement~\cite{:2010px}).
To establish baseline behavior, ATLAS has used the spectrum of single muons
at large $\pT$ to extract the yield of $W^{\pm}$ bosons, by a
simultaneous fit of a template trained on simulated $W^\pm$ as well as 
a parameterization of muons from heavy flavor quark decays, 
as shown in the left panel of Fig.~\ref{W}.
A sample of roughly 400 $W$ bosons has been extracted. 
The binary scaling of the measured yields is studied using the variable 
$R_{\mathrm{PC}}$, defined as the ratio of yields measured in different 
centrality classes to the yield measured in the 10\% most central events,
with all yields scaled by the corresponding number of binary nucleon-nucleon collisions.
The right panel of Fig.~\ref{W} shows $R_{\mathrm{PC}}$ as a function of centrality~\cite{QM2011W}.
Using a fit to a constant value, which gives $\langle R_{\mathrm{PC}} \rangle$ = 0.99$\pm$0.10 
with a $\chi^2=3.02$ for 3 degrees of freedom, 
a significant consistency with binary scaling is observed.


\begin{figure}[t]
\begin{center}
\includegraphics[width=16pc]{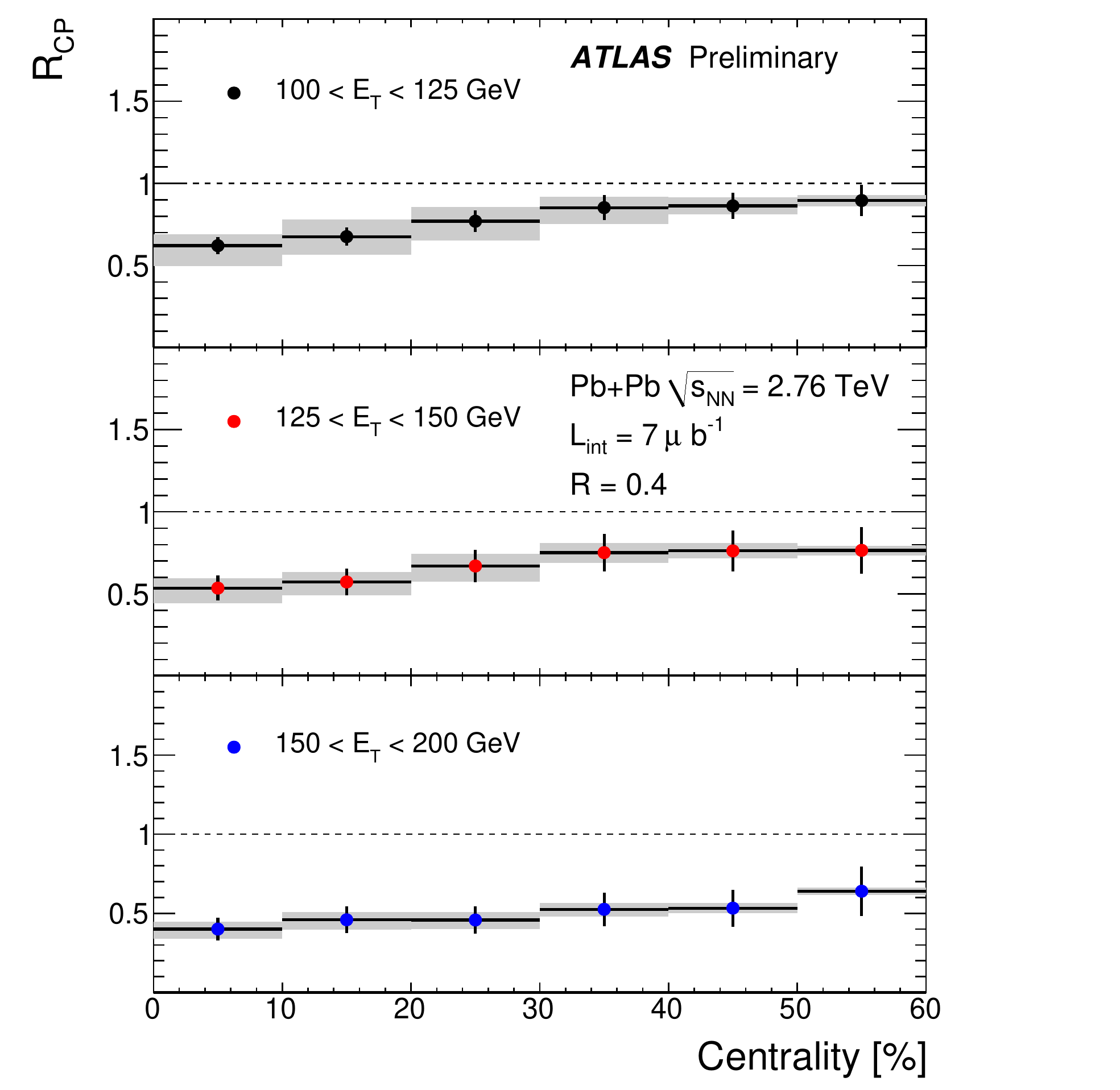}
\includegraphics[width=16pc]{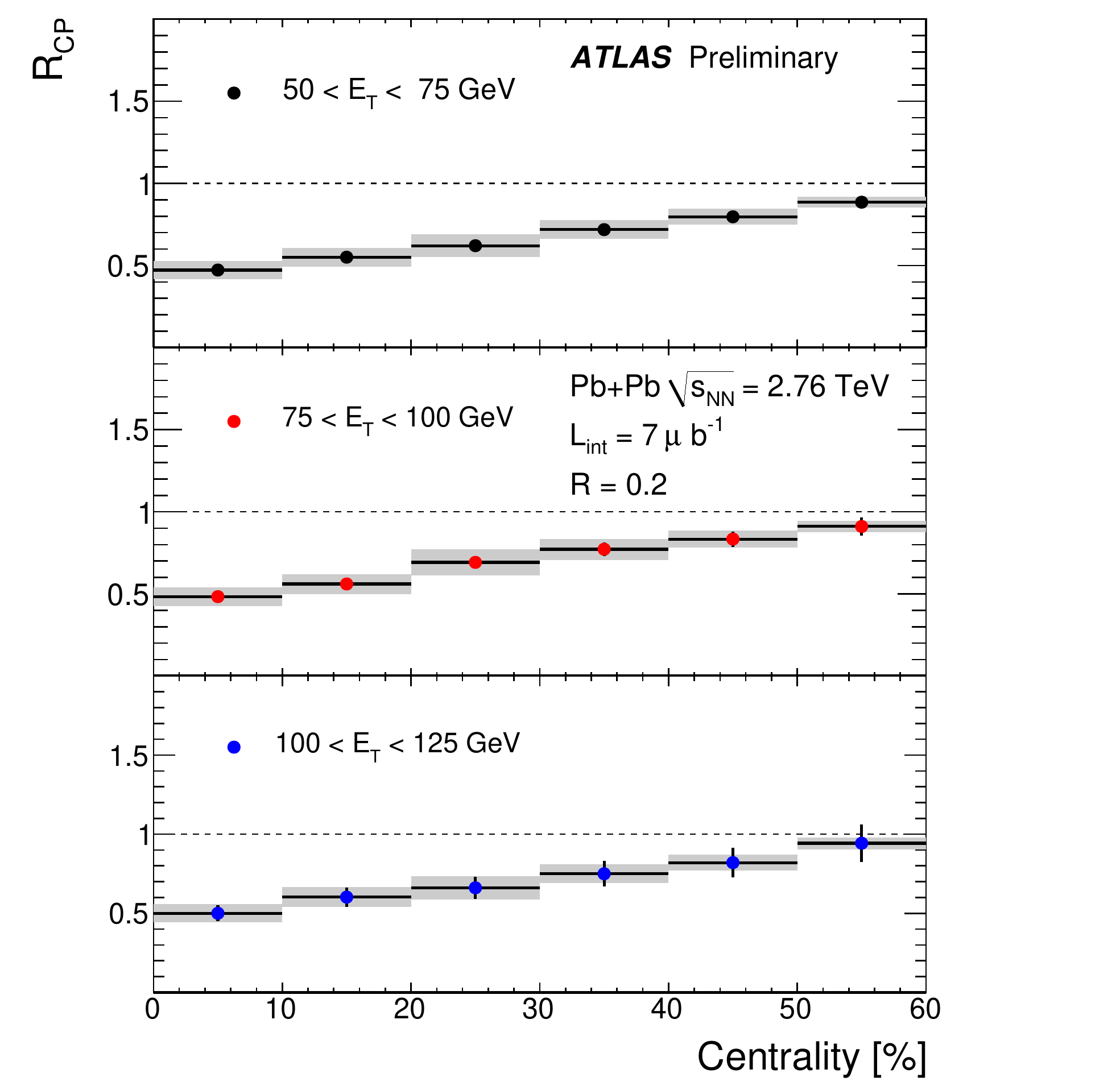}
\caption{\label{jet_rcp}
$R_{CP}$ for R=0.4 (left) and R=0.2 (right) jets, as a function of centrality,
with 60-80\% used as the peripheral sample.  In this convention, the most
central events are on the left and the most peripheral are on the right.
}
\end{center}
\end{figure}

\begin{figure}[t]
\begin{center}
\includegraphics[width=16pc]{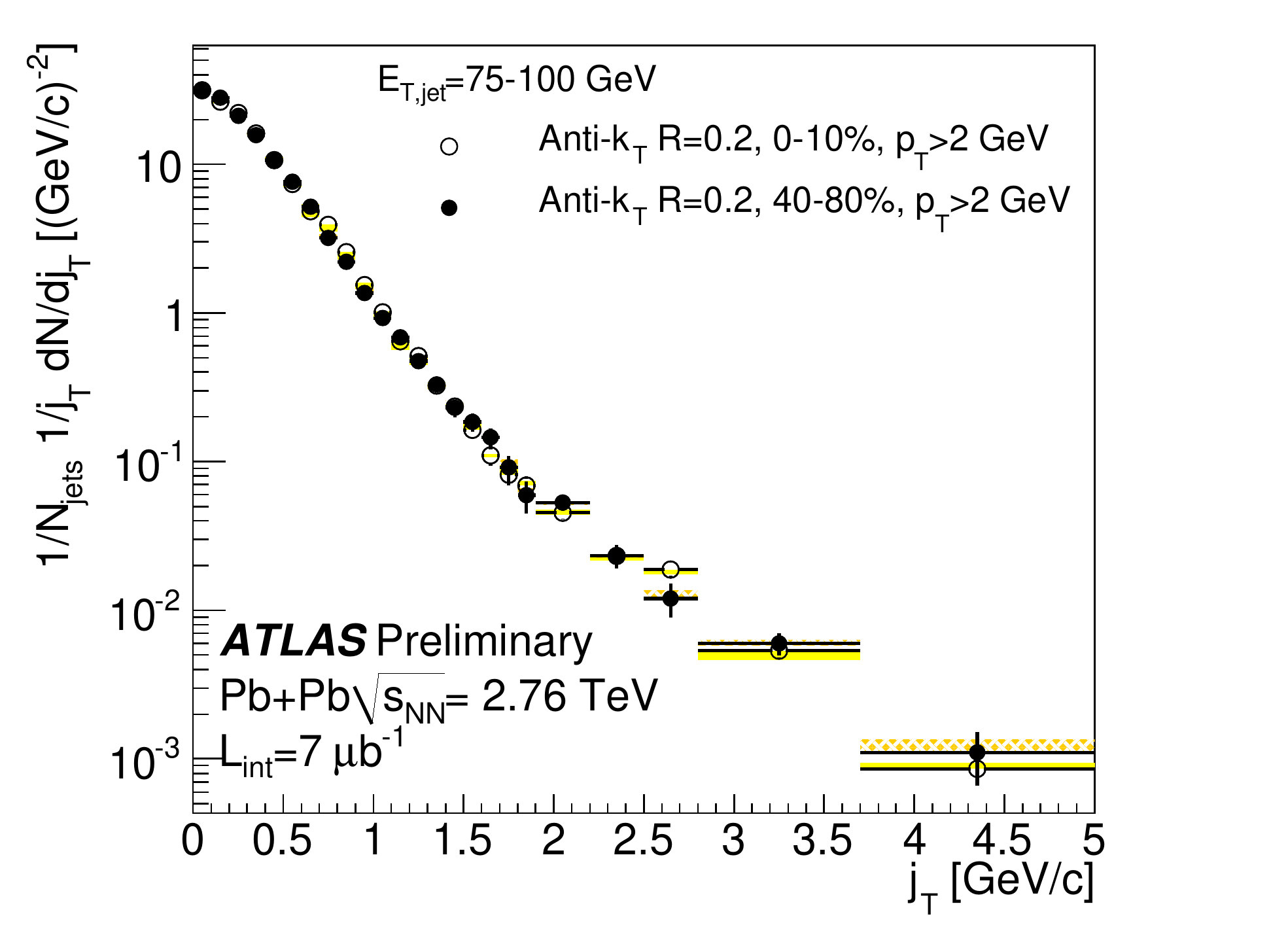}
\includegraphics[width=16pc]{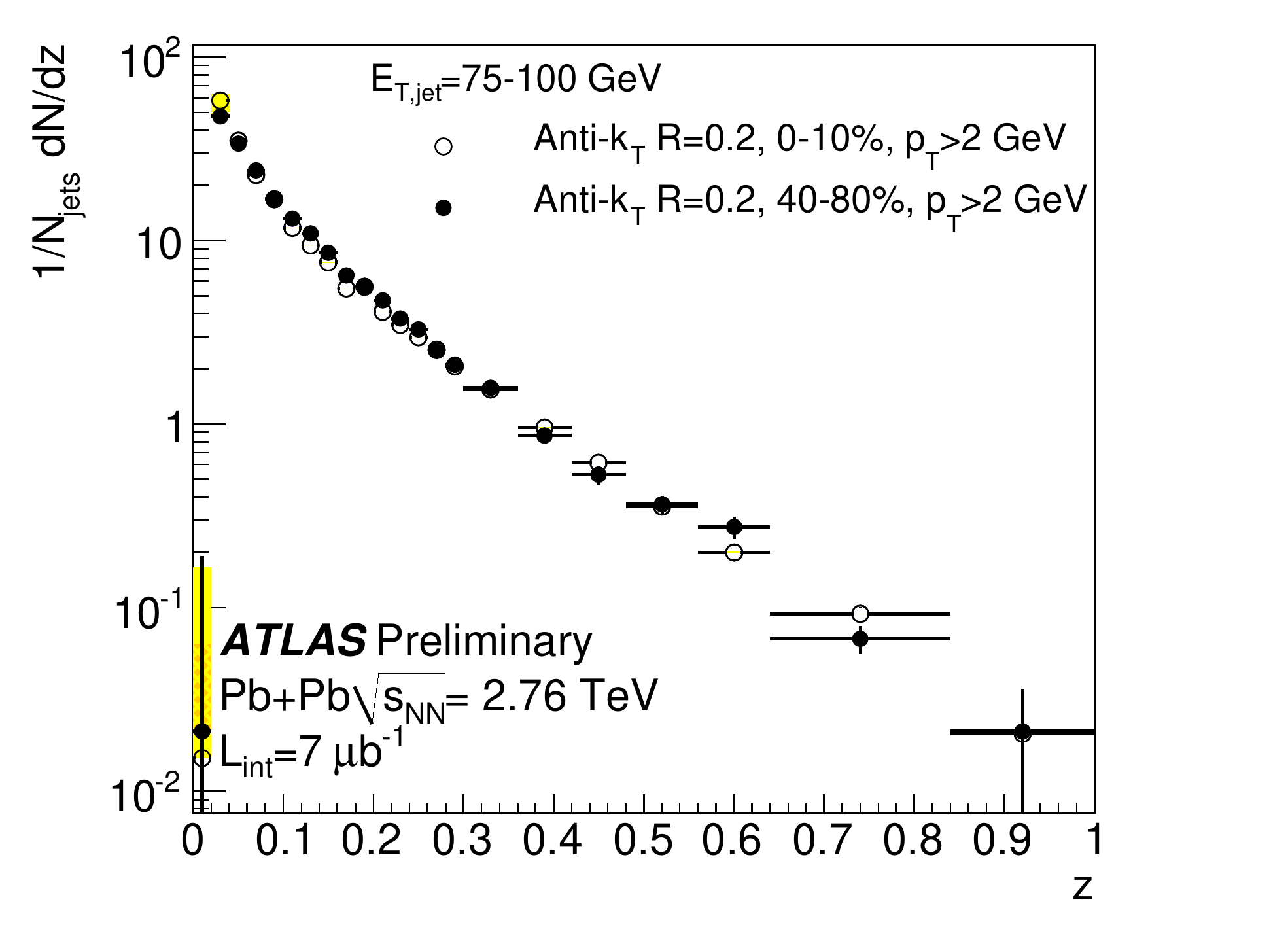}
\caption{\label{frag_func}
(left) Transverse fragmentation function for R=0.2 anti-$k_t$ jets, comparing central 0-10\% events with peripheral 40-80\%.
(right) the similar comparison for the longitudinal fragmentation function.
}
\end{center}
\end{figure}

ATLAS has already published striking results on dijet asymmetries~\cite{ATLASjetq}.
New results are reported here on the
measurement of inclusive jet yields as a function
of jet $\eT$, in order to test models of QCD energy loss more directly~\cite{QM2011jets}.
Jets are reconstructed  using the anti-$k_t$ algorithm
with the jet size set to both R=0.2 and R=0.4, based on ``towers'' composed of
calorimeter cells integrated over regions of size $\Delta\eta \times \Delta\phi = 0.1\times 0.1$.
The ambient background is removed at the cell level by excluding regions 
near jets and calculating the mean energy, as well as any azimuthal modulation,
in strips of width $\Delta\eta=0.1$.
An iterative procedure is applied to remove any residual effect of the jets
on the background subtraction.
The final jets are corrected for jet energy scale
and resolution based on PYTHIA jets embedded into HIJING,
with conservative systematic uncertainties assigned to account for the
small differences between the fluctuations seen in data and simulation.
Jets are restricted to $|\eta|<2.8$, to stay within
the main barrel and endcap regions of the calorimeter.

Figure~\ref{jet_rcp} shows $R_{\mathrm{CP}}$ (defined similarly to $R_{\mathrm{PC}}$ bin,
but defined relative to the 60-80\% centrality interval) as a function of centrality for jets
in three fixed $\eT$ bins ($\eT = 100-125$ GeV, $125-150$ GeV and $150-200$ GeV).
A smooth evolution of $R_{\mathrm{CP}}$ is observed going from the 50-60\% central events
on the right to the 0-10\% central events on the left, where a suppression of
roughly a factor of two is observed for all $\eT$ bins and for both R=0.4
as well as R=0.2.  The invariance with R is suprising given the general 
tendency of radiative calculations to predict a modified fragmentation functions for
quenched jets.
However, a direct measurement of longitudinal and transverse fragmentation functions, shown
in Fig.~\ref{frag_func}, confirms that no substantial modification of the
fragmentation function can be observed comparing peripheral and central events.
In other words, the increased suppression of the jet rates
is not accompanied by any evident modification of the jets themselves.

\begin{figure}[t]
\begin{center}
\includegraphics[width=24pc]{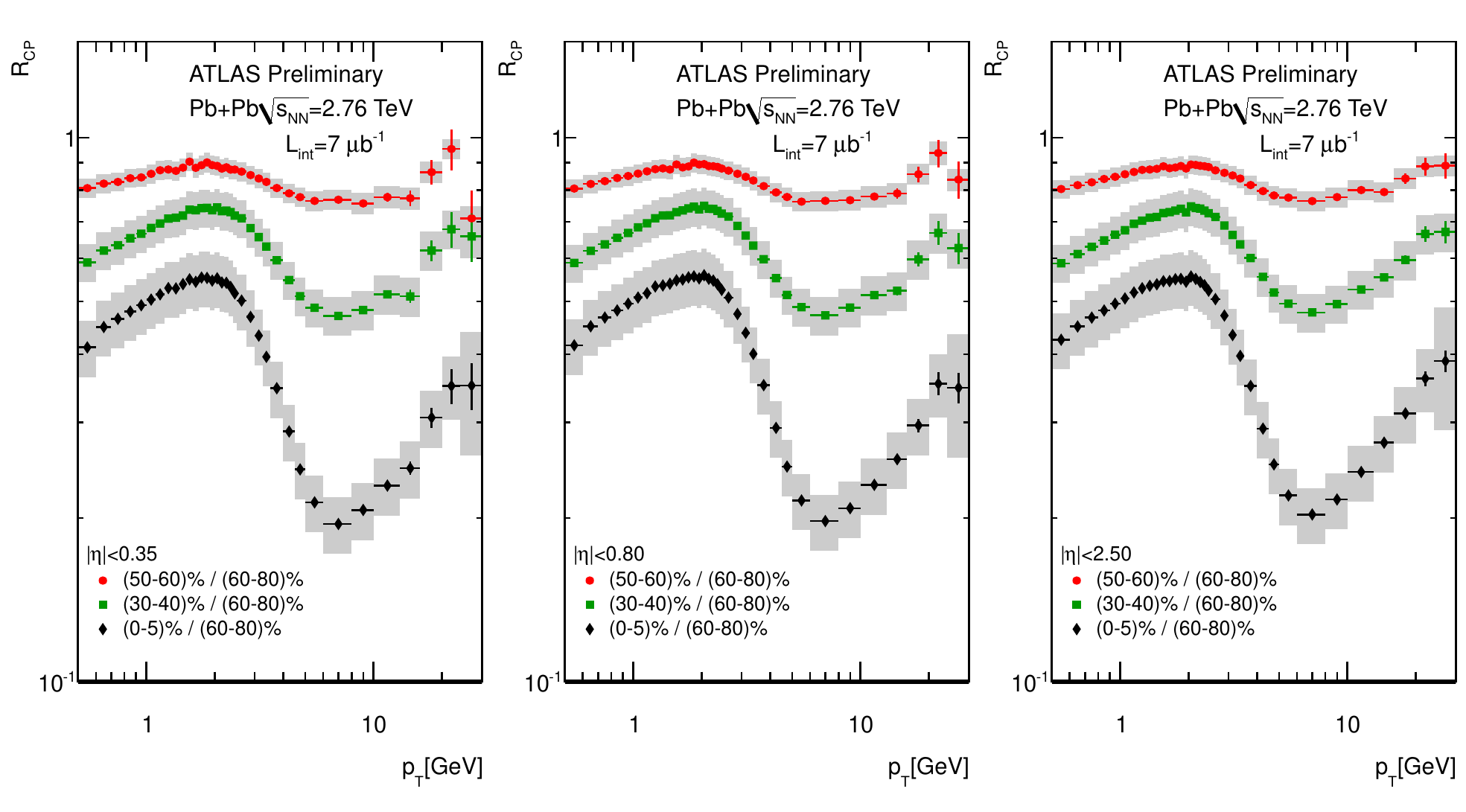}
\includegraphics[width=18pc]{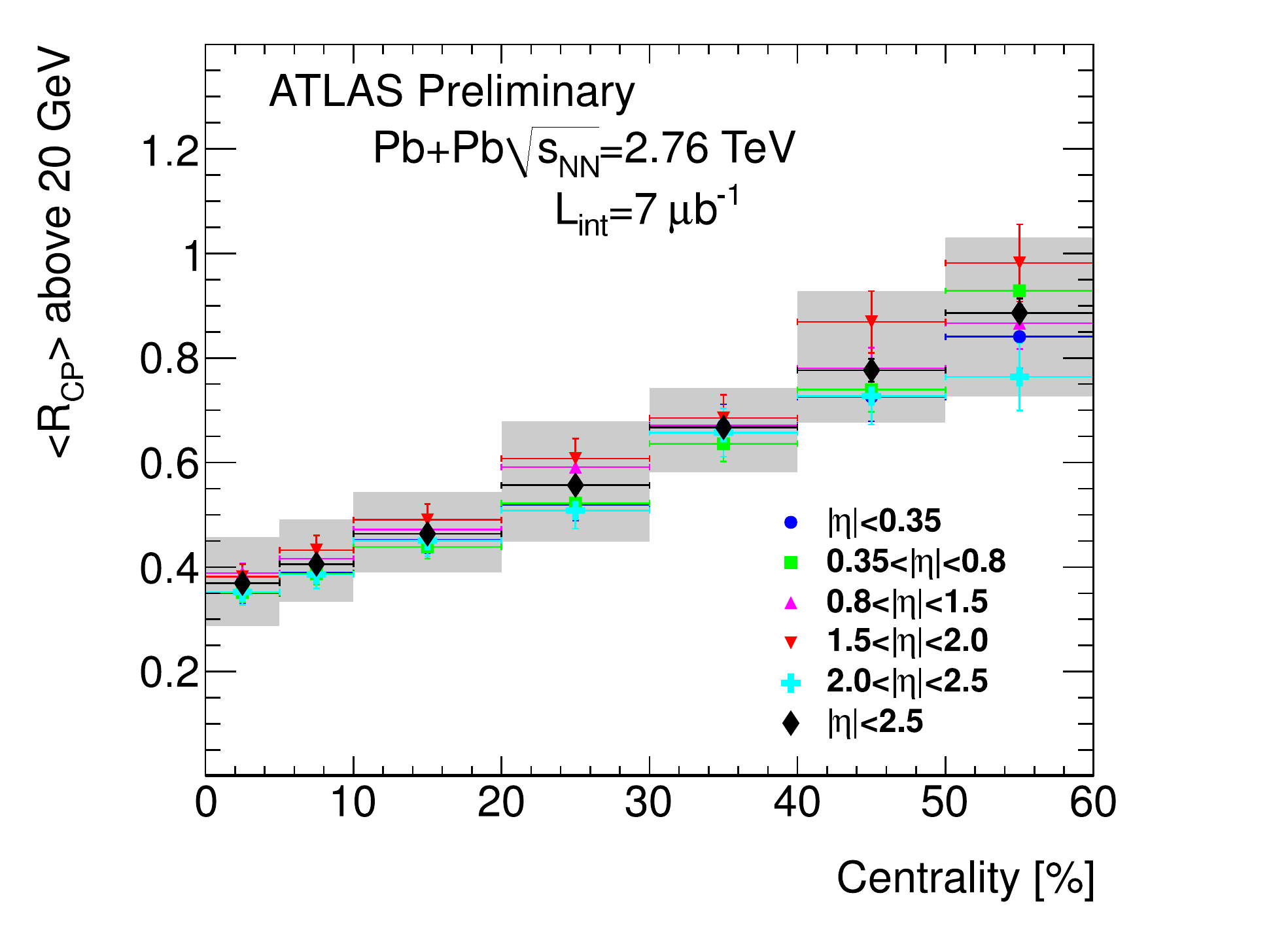}
\caption{\label{charged_rcp}
(top) Charged particle $R_{\mathrm{CP}}$ measured relative to the 60-80\% bin, for all
tracks measured within $|\eta|<0.35$, $|\eta|<0.8$ and $|\eta|<2.5$.  
(bottom) $R_{\mathrm{CP}}$ for $p_T = 20-30$ GeV as a function of centrality.
}
\end{center}
\end{figure}

To study this phenomenon further, charged particle spectra have been measured
out to $|\eta| = 2.5$ and $\pT = 30$ GeV~\cite{QM2011spec}.  Tracks are reconstructed in
the full Inner Detector, with stringent requirements on the number of hits
per track in order to suppress fake tracks.  With these requirements,
efficiencies are typically 70-80\% while fake rates near $\eta=0$ are
less than a percent, although both efficiency and purity are reduced somewhat
going to forward rapidity.  While tracks are measured in ATLAS at transverse momenta
well past 30 GeV, the measured track impact parameter errors (which are also used to reject
fake tracks) are not well described by the full ATLAS simulation.
Figure~\ref{charged_rcp} top shows $R_{CP}$ measured for charged
hadrons relative to the 60-80\% centrality interval.  The trend is similar
to that seen in charged particle $R_{AA}$ measured by ALICE~\cite{Aamodt:2010jd}.
There is a peak at 2-3 GeV, a minimum around 7 GeV and then a
rise which has not yet leveled off even at $\pT = 30$ GeV.
The value of $R_{CP}$ averaged over $\pT$ from 20 to 30 GeV, shown in the bottom panel of Fig.~\ref{charged_rcp}, is found to decrease smoothly 
from a value near unity for 50-60\% to approximately 0.4 for the 0-10\% most central collisions, 
a trend both qualititatively and quantitatively similar to that found
for fully reconstructed jets.

\section{Conclusions}

Results are presented from the ATLAS detector from the 2010 LHC heavy ion run.
The charged particle multiplicity
increases by a factor of two relative to the top RHIC energy, with a 
centrality dependence very similar to that already measured at RHIC.
Measurements of elliptic flow out to large transverse
momentum also show similar results to what was measured at RHIC,
but no significant pseudorapidity dependence.
Measurements of higher harmonics have also been made,
which seem to explain structures in the two-particle correlation
previously attributed to jet-medium interactions.
Single muons at high momentum are used to extract the yield of $W^{\pm}$ bosons
and are found to be consistent with binary collision scaling.
Conversely, jets are found to be suppressed in central events by a factor of
two relative to peripheral events, 
with no significant dependence on the jet energy.
Fragmentation functions
are also found to be the same in central and peripheral events.
Finally, charged hadrons have been measured out to 30 GeV, and are found to have a centrality
dependence (relative to peripheral events) similar to that found for jets.


\section*{References}
\begin{thebibliography}{10}
\bibitem{Aad:2008zzm} ATLAS Collaboration, G.~Aad et al., JINST {\bf 3}, S08003 (2008).
\bibitem{QM2011Mult} Y.~Chen, these proceedings.  ATLAS Collaboration, ``Measurement of the centrality dependence of the charged-particle pseudorapidity distribution in lead-lead collisions at $\sqrt{s_{NN}} = 2.76$~TeV with the ATLAS detector,'' to be submitted to Phys. Lett. B.
\bibitem{Collaboration:2010cz} ALICE Collaboration, K.~Aamodt  {\it et al.}, Phys.\ Rev.\ Lett.\  {\bf 106} (2011) 032301.
\bibitem{QM2011flow} A.~Trzupek, these proceedings.  J.~Jia, these proceedings.  ATLAS Collaboration, ``Pseudorapidity and transverse momentum dependence of elliptic flow for charged particles in lead-lead collisions at $\sqrt{s_{_{NN}}}=2.76$ TeV with the ATLAS detector at the LHC,'' to be submitted to Phys.~Lett.~B.  ATLAS-CONF-2011-074 (\url{http://cdsweb.cern.ch/record/1352458}.)
\bibitem{alicepaper} {K.~Aamodt et al., ALICE Collaboration, Phys. Rev. Lett. 105 (2010) 252302.}
\bibitem{Adare:2010sp} A.~Adare {\it et al.} PHENIX Collaboration,  Phys.\ Rev.\ Lett.  105 (2010) 142301.
\bibitem{Adams:2004bi} J.~Adams {\it et al.}, STAR Collaboration, Phys.\ Rev.\  C 72  (2005)  014904.

\bibitem{Alver:2010gr} B.~Alver and G.~Roland,  Phys.\ Rev.\  C {\bf 81}, 054905 (2010)  [Erratum-ibid.\  C {\bf 82}, 039903 (2010)].
\bibitem{CasalderreySolana:2004qm} J.~Casalderrey-Solana, E.~V.~Shuryak, D.~Teaney,  J.\ Phys.\ Conf.\ Ser.\  {\bf 27}, 22-31 (2005).
\bibitem{Adams:2005ph} J.~Adams {\it et al.},  Phys.\ Rev.\ Lett.\  {\bf 95} (2005) 152301.
\bibitem{Adare:2007vu} A.~Adare {\it et al.},  Phys.\ Rev.\  C {\bf 77} (2008) 011901.
\bibitem{:2010px} The ATLAS Collaboration, G.~Aad {\it et al.}, Phys.\ Lett.\  {\bf B697} (2011) 294.
\bibitem{QM2011W} R.~Sandstr\"{o}m, these proceedings.  ATLAS-CONF-2011-078(\url{http://cdsweb.cern.ch/record/1353227}.)
\bibitem{ATLASjetq} ATLAS Collaboration,  G.~Aad  {\it et al.},  Phys.\ Rev.\ Lett.\  {\bf 105} (2010) 252303.
\bibitem{QM2011jets} B.~Cole, these proceedings.  A.~Angerami, these proceedings.  ATLAS-CONF-2011-075 (\url{http://cdsweb.cern.ch/record/1353220}.)
\bibitem{QM2011spec} A.~Milov, these proceedings.  ATLAS-CONF-2011-079 (\url{http://cdsweb.cern.ch/record/1355702}.)
\bibitem{Aamodt:2010jd} K.~Aamodt {\it et al.},  Phys.\ Lett.\  B {\bf 696} (2011) 30.
\end{thebibliography}
\end{document}